\documentclass{JHEP3}
 \usepackage{epsfig}

 \title{
  The underlying event and fragmentation
  }

 \author{Kosuke Odagiri\\
  Institute of Physics, Academia Sinica, Nankang, Taipei, Taiwan 11529,
  The Republic of China}

 \abstract{
  A good fit to the CDF underlying event is obtained in the multiple
parton scattering picture using HERWIG, after modifying the cluster
hadronization algorithm as suggested by our previous study and adopting a
larger maximum cluster size.
  The number of scatters per event is generated simply as a Poisson
distribution.
  If our picture is correct, the baryon yield should be enhanced in the
underlying event. This effect may be studied by measuring the
proton-to-pion ratio.
  }


 \preprint{hep-ph/0407008}

\begin{document}

 \section{Introduction}

  The quantitative description of soft multiple-parton interaction in
hadronic collision has been a long-standing and unsolved problem.
  Such interaction, which for instance contributes to the underlying 
hadronic activity, the `underlying event', has received much attention 
recently, both in terms of experiment at Tevatron \cite{chargedjet} and 
theory 
\cite{sjostvanzijl,Marchesini:1988hj,jimmy,borozan,Sjostrand:2004pf}.
  With increased collision centre-of-mass energy, one expects a 
proliferation of such hadronic activity. The current operation of 
Tevatron, as well as the operation of LHC being in plan for the near 
future, adds urgency to its study.

  In terms of experiment, in the simplest language, the underlying event
appears as a uniform plateau of hadronic transverse energy which must be
subtracted from jets.
  The uncertainty in this jet energy correction is the dominant source of
the systematic error on jet energy at low transverse momentum even at CDF.
The effect is expected to become more significant
\cite{sjostvanzijl,Marchesini:1988hj} with increased
centre-of-mass-energy, such as at LHC.

  There has recently been significant progress \cite{chargedjet} in the
measurement of hadronic activity in limited regions of phase space that
are defined with respect to the direction of the leading jet. These
results have been fitted with the available Monte Carlo event generators
\cite{herwig,pythia} with some success.

  Another notable measurement has been that of the rate of double parton
scattering in the process $p\bar p\to \gamma/\pi^0+3\mathrm{j}+X$ at CDF
\cite{doublescat}. Their measurement gave a surprisingly small value for
the `effective cross section'
$\sigma_\mathrm{eff}=14.5\pm1.7^{+1.7}_{-2.3}$ mb which, however, seems
not to have been taken account in many theoretical studies since then.

  Progress has been made on the theoretical side mainly by considering
multiple $(2\to2)$ scatters whose frequency is governed by a given
`overlap function'. Events are then generated by coupling the $(2\to2)$
parton-level scatters with some preferred Monte Carlo event generator.

  The superficial convergence of the theoretical effort, which might give
one the impression that all that is left is to tune the Monte Carlo event
generators, is misleading. The following are a few theoretical reasons as
to why this is the case.

  First, the fragmentation of low-transverse-momentum (low-$p_T$) partons
is not well-understood. Monte Carlo event generators are tuned to describe
data from, for instance, LEP \cite{tunes}. Here the usual prescription of
perturbative parton shower up to a cut-off followed by hadronization, and
hence the Monte Carlo event generators, work very well. This is because
the distribution of hadrons is governed primarily by the partons emitted
in the parton shower phase and this phase is well-understood. What happens
in the low-$p_T$ case where there is potentially little perturbative
emission is far from clear.

  Second, the above `overlap function' approach is based on the assumption
that the proton is a `disk' with a radius-dependent density, and partons
within the proton interact independently of each other. This assumption
needs justification.
  In particular, when the interaction distance becomes comparable or
larger than the proton `radius', this approach would seem insufficient.  
Even for a process containing a high $p_T$ subprocess, the lowest energy
component of the process occurs near the QCD scale, which is comparable
with the proton radius.

  Third, the mean number of scatters is proportional to the inclusive
cross section which is critically dependent on the $p_T$ cut-off
$p_T^\mathrm{min}$. At the moment, this imposition of $p_T^\mathrm{min}$
is an ad-hoc procedure and we need to obtain more understanding of the
behaviour of the low $p_T$ scatters, both in terms of the inclusive cross
section and the fragmentation properties.

  In addition, there are some phenomenological insufficiencies which are
possibly related to the above theoretical problems.
  We merely outline the problems here. A more detailed discussion will be
provided later on.

  First, the mean number of scatters is inversely proportional to the
above-mentioned `effective cross section'. The overlap function approach,
if based on the low-energy proton form factor, predicts effective cross
sections that are much larger than the measured value.

  Second, the predicted underlying event is too soft, i.e., the hadrons 
carry too little transverse momentum.

  Third, the predicted underlying event is too uniform.

  The purpose of this paper is to propose a step towards the solution to
the above problems within the framework of the multiple-scattering
picture.

  We first adopt the measured value of $\sigma_\mathrm{eff}$ and set
$p_T^\mathrm{min}$ near the Regge-perturbative `transition' scale so that,
to a good approximation, each scatter can be calculated using perturbation
theory. The number of scatters is generated according to the Poisson
distribution.
  We find that the level of underlying event activity is in agreement with
the experimental findings.

  We then study the fragmentation scheme dependence using the HERWIG
Monte-Carlo event generator \cite{herwig}.

  In a recent study \cite{odagirihadronization}, we suggested that the
cluster splitting process in the cluster fragmentation model
\cite{cluster_hadronization} of HERWIG may be reducible to a modified
low-energy $\alpha_S$ that describes `unresolved' emission below the
parton shower cut-off.
  In accord with the expectation from this study, the shape of the
measured underlying event is described well by adopting the modified
hadronization scheme and a larger cluster mass cut-off.

  This paper is organized as follows. In sec.~\ref{sec_nature}, we discuss 
the current status of theoretical and experimental study, and outline our 
interpretation of the discrepancies. In sec.~\ref{sec_simulation}, we 
present and discuss the result of our simulation using HERWIG. The 
conclusions, and the outlook regards future application, follow in 
sec.~\ref{sec_conclusions}.

 \section{Nature of interaction}\label{sec_nature}

 \subsection{General remarks}

  Multiple-parton interaction is often classified as a unitarization
effect.
  The total inclusive jet cross section is defined experimentally by
counting the number of jet pairs rather than the number of events, and
theoretically by the ununitarized perturbative calculation for a given
minimum transverse momentum $p_T^\mathrm{min}$.
  For sufficiently large centre-of-mass energy $\sqrt{s_\mathrm{CM}}$ and
small $p_T^\mathrm{min}$, the inclusive jet cross section, either
calculated or measured, in general exceeds the relevant total cross
section. The scattering probability needs to be unitarized so that per
event, the mean number of scatters $\overline n$ of a `type' of
interaction that has inclusive cross section $\sigma^\mathrm{inc}$ and
total cross section $\sigma^\mathrm{tot}$ is given by:
 \begin{equation}
  \overline n=\frac{\sigma^\mathrm{inc}}{\sigma^\mathrm{tot}}.
  \label{eqnnbar}
 \end{equation}

  Let us now consider, according to some suitable definition, the whole of
non-diffractive (ND) events. In an ND event we then have on average
$\overline n_A=\sigma^\mathrm{inc}_A/\sigma^\mathrm{tot}_\mathrm{ND}$
scatters of the type $A$.

 \subsection{The Poisson approach}

  The distribution of the number of scatters is not uniquely determined.  
The simplest approach is to assume that the scatters are, neglecting the
possible restriction coming from the conservation of energy which turns
out not to be a significant effect, independent of each other. We then
have a Poisson distribution of the number of scatters. The normalization
is such that the cross section for events containing $n\geq1$ scatters is
given by:
 \begin{equation}
  \sigma_n=\frac{\nu^{n-1}}{n!}e^{-\nu}\sigma^\mathrm{inc}_\mathrm{ND},
  \label{eqnsigind}
 \end{equation}
  $\nu$ being a constant.
  The inclusive and total non-diffractive cross sections are then related
by:
 \begin{eqnarray}
  \sigma^\mathrm{inc}_\mathrm{ND}&=&\sum_{n\geq1}n\sigma_n,\\
  \sigma^\mathrm{tot}_\mathrm{ND}&=&\sum_{n\geq1}\sigma_n
   =\frac{1-e^{-\nu}}\nu\sigma^\mathrm{inc}_\mathrm{ND},
 \end{eqnarray}
  so that:
 \begin{equation}
  \overline n =\frac{\sigma^\mathrm{inc}_\mathrm{ND}}
    {\sigma^\mathrm{tot}_\mathrm{ND}}
   =\frac\nu{1-e^{-\nu}}.
 \end{equation}
  $\overline n\to\nu$ in the limit of large $\nu$. Now let us define the
following quantity, applicable to this case, which matches with the
`effective cross section' of ref.~\cite{doublescat}:
 \begin{equation}
  \sigma_\mathrm{eff}=\frac{\sigma^\mathrm{tot}_\mathrm{ND}}
  {1-e^{-\nu}}>\sigma^\mathrm{tot}_\mathrm{ND}.
 \end{equation}

  In ref.~\cite{doublescat}, $\sigma_\mathrm{eff}$ was measured for the
1.8 TeV run at CDF by comparing the rate of back-to-back
jet+$\gamma/\pi^0$ events that also contain a back-to-back dijet pair,
against similar events due to pile-up. Systematic uncertainties were hence
reduced compared with previous studies and they obtain:
 \begin{equation}
  \sigma_\mathrm{eff}=14.5\pm1.7_{-2.3}^{+1.7} \ \mathrm{mb}.
  \label{eqnsigeff}
 \end{equation}
  We note that this number is small compared with the measured value
\cite{cdftotal} of the nondiffractive total cross section at CDF. At 1.8
TeV they find:
 \begin{eqnarray}
  \sigma_\mathrm{NSD}^\mathrm{total}&=&
  \sigma^\mathrm{total}_\mathrm{inelastic}-
  \sigma^\mathrm{total}_\mathrm{SD} \nonumber\\
  &=& (60.33\pm1.40)-(9.46\pm0.44) = 50.87\pm1.5 \ \mathrm{mb}.
  \label{eqnsignsd}
 \end{eqnarray}
  Here SD stands for `single diffractive' events where one of the protons 
remains intact. NSD stands for non-SD.

  The surprisingly large difference between the two cross sections 
$\sigma_\mathrm{eff}$ and $\sigma_\mathrm{NSD}^\mathrm{total}$ can be 
accommodated within the above naive Poisson framework if the remaining 
35~mb of the NSD cross section is in fact due to the exchange of 
colour-singlet objects, the Pomeron and Reggeons, and does not resolve the 
quark structure of the proton.
  If we adopt this viewpoint, the `non-diffractive' cross section should
perhaps be understood rather as a `resolved' cross section in analogy with
the resolved photon cross section. Later on in this paper, we use the
notation $\sigma_\mathrm{res}$ to indicate this cross section.

 \subsection{The overlap-function approach}

  A commonly adopted approach in parametrizing the underlying events is to
modify the picture of independent scatters by introducing an `overlap
function'. In this picture, an event occurs at a definite impact parameter
$b$, and scatters are independent per event, i.e., eqn.~(\ref{eqnsigind})
is replaced by:
 \begin{equation}
  \sigma_n=\int_0^\infty\frac{\nu^n(b)}{n!}e^{-\nu(b)}\pi db^2.
  \label{eqnsigovlp}
 \end{equation}
  The inclusive and total cross sections are now given by:
 \begin{eqnarray}
  \sigma^\mathrm{inc}_\mathrm{ND}&=&\sum_{n\geq1}n\sigma_n
     = \int_0^\infty \nu(b)\pi db^2, \label{eqnsigincovlp}\\
  \sigma^\mathrm{tot}_\mathrm{ND}&=&\sum_{n\geq1}\sigma_n
     = \int_0^\infty\left[1-e^{-\nu(b)}\right]\pi db^2.
 \end{eqnarray}
  After this, one normally adopts a factorized form for $\nu$, namely:
 \begin{equation}
  \nu(b)=\sigma^\mathrm{inc}_\mathrm{ND}A(b).
 \end{equation}
  $A(b)$ is the overlap function which, from eqn.~(\ref{eqnsigincovlp}) 
satisfies the normalization condition:
 \begin{equation}
  \int_0^\infty A(b)\pi db^2=1.
 \end{equation}
  We can evaluate $\sigma_\mathrm{eff}$ from $A(b)$. For two
distinguishable scatters $A$ and $B$ with small cross sections,
i.e., $\sigma_A,\sigma_B$ being much less than
$\sigma^\mathrm{tot}_\mathrm{ND}$, we have:
 \begin{equation}
  \sigma_\mathrm{eff}=\frac{\sigma_A\sigma_B}{\sigma_{AB}}
   =\left[\int_0^\infty A^2(b)\pi db^2\right]^{-1}.
  \label{eqnsigeffovlp}
 \end{equation}

  Here we are led into a dilemma. If we are to use a physically motivated
overlap function based on the proton form factor, the resulting cross
section is too large. For instance, in refs.~\cite{jimmy,borozan} we have:
 \begin{equation}
  A(b)=\frac{\mu^2}{96\pi}(\mu b)^3K_3(\mu b),
 \end{equation}
  where $\mu^2=0.71$ GeV$^2$ and $K_i(x)$ are the modified Bessel
functions. Substituting this into eqn.~(\ref{eqnsigeffovlp}), we obtain
$28\pi/\mu^2\approx48$ mb, which is too large.

  We add that in terms of theory, the classical picture of parton-filled
protons interacting according to the parton density given by some form
factor does not have strong justification. In particular, the typical soft
QCD interaction distance is of order of, or even larger than, the proton
radius. Even events with the hard subprocess at scales much greater than
the QCD scale in general contain a usually factorizable soft part.

  Let us consider how the simple Poisson distribution of
eqn.~(\ref{eqnsigind}) differs from the case with a varying overlap
function, eqn.~(\ref{eqnsigovlp}). A simple test case is that of the
Gaussian overlap function:
 \begin{equation}
  \nu(b)=\frac{\lambda\sigma^\mathrm{tot}}{\pi}e^{-\lambda b^2}.
 \end{equation}
  $\lambda$ is a dimension-2 constant.
  With this choice of $\nu(b)$, eqn.~(\ref{eqnsigind}) can be evaluated
analytically, and we obtain:
 \begin{eqnarray}
  \sigma_n&=&\frac{\pi}{n\lambda}
  \left[1-e^{-\lambda\sigma^\mathrm{inc}/\pi}
  \sum_{m=0}^{n-1}\frac1{m!}({\lambda\sigma^\mathrm{inc}/\pi})^m
  \right]\\
  &\equiv&\frac{\pi}{n\lambda}
  e^{-\lambda\sigma^\mathrm{inc}/\pi}
  \sum_{m=n}^\infty\frac1{m!}({\lambda\sigma^\mathrm{inc}/\pi})^m.
  \label{eqnsiggauss}
 \end{eqnarray}
  The inclusive cross section satisfies:
 \begin{equation}
  \sigma^\mathrm{inc}=\sum_{n=1}^\infty n\sigma_n,
 \end{equation}
  whereas the total cross section is given by:
 \begin{eqnarray}
  \sigma^\mathrm{tot}&=&\sum_{n=1}^\infty\sigma_n\nonumber\\
  &=&\int_0^\infty\pi db^2\left[1-e^{-\nu(b^2)}\right]\nonumber\\
  &=&\frac\pi\lambda\int_0^{\lambda\sigma^\mathrm{inc}/\pi}
  \frac{d\nu}\nu\left[1-e^{-\nu}\right].
 \end{eqnarray}
  Hence the mean number of scatters per event is:
 \begin{equation}
  \overline{n}=\frac{\sigma^\mathrm{inc}}{\sigma^\mathrm{tot}}
  \approx 1+\frac{\lambda\sigma^\mathrm{inc}}{4\pi}
 \end{equation}
  This last approximation is accurate up to about $\left<n\right>=10$.

  The largest terms in the series expansion of the exponential are found
at the order that is close to the expansion coefficient. We therefore see
that for small $n$, eqn.~(\ref{eqnsiggauss}) has an approximate $1/n$
behaviour. At large $n$, this is replaced by a Poisson-like behaviour.  
$\sigma_n$ decreases monotonically with $n$, in contrast with the Poisson
case where $\sigma_n$ is largest at $n=\overline{n}$.

  Thus the difference between the two approaches is in principle
significant.

  In any case, the deficiencies of the overlap function approach
highlights the need to rethink the low-energy behaviour of the scattering
cross sections.

  The mean number of scatters per event is not too large at Tevatron for
reasonable $p_T^\mathrm{min}$ and the measured value of
$\sigma_\mathrm{eff}$. For instance, in ref.~\cite{borozan}, the mean
number of scatters per event is 2.4. Hence the effect of adopting
different distributions is expected not to be drastic. On the other hand,
for larger centre-of-mass energy, for instance at LHC, such effect may
become significant. This should be a topic for future experimental
studies.

  In addition, the measurement of $\sigma_\mathrm{eff}$ on distinct
experimental platforms, such as $\gamma\gamma$ and $\gamma p$ colliders,
would be an interesting possibility.

 \subsection{The CDF `charged jet' analysis}

  In the analysis of ref.~\cite{chargedjet}, hadronic activity in the
underlying event is studied with respect to the leading `charged jet',
which is defined using a non-standard cone algorithm, counting only the
charged tracks with $p_T$ greater than 0.5 GeV and pseudorapidity $\eta$
between $-1$ and $+1$.
  Jet $p_T$ is defined as the scalar sum of the $p_T$ of the charged 
tracks within the jet.

  The charged jet construction algorithm proceeds as follows.
  First, charged tracks are ordered in decreasing $p_T$.
  One then starts with the highest $p_T$ track, and define this as the
seed to form a jet.
  Going down the list of the remaining charged tracks, tracks which are
less than $\Delta R<0.7$ away from this seed are combined with it. Here
$\Delta R=\sqrt{\Delta\eta^2+\Delta\phi^2}$ as usual, with $\phi$ being
the azimuthal angle.
  The new seed is defined to be at the centroid of the two objects. This
is defined in the $\eta-\phi$ space, using $p_T$ as the weight. The $p_T$
of the new object is the scalar sum of the $p_T$ of the two constituents.
  This process is repeated until the list is exhausted, after which one
starts again using the highest $p_T$ track that remains.
  After all charged tracks have been assigned to jets, with the
possibility that some jets may contain just one track, the jet
construction algorithm is terminated, and the leading charged jet is
defined as the jet with the largest $p_T$.

 \EPSFIGURE[ht]{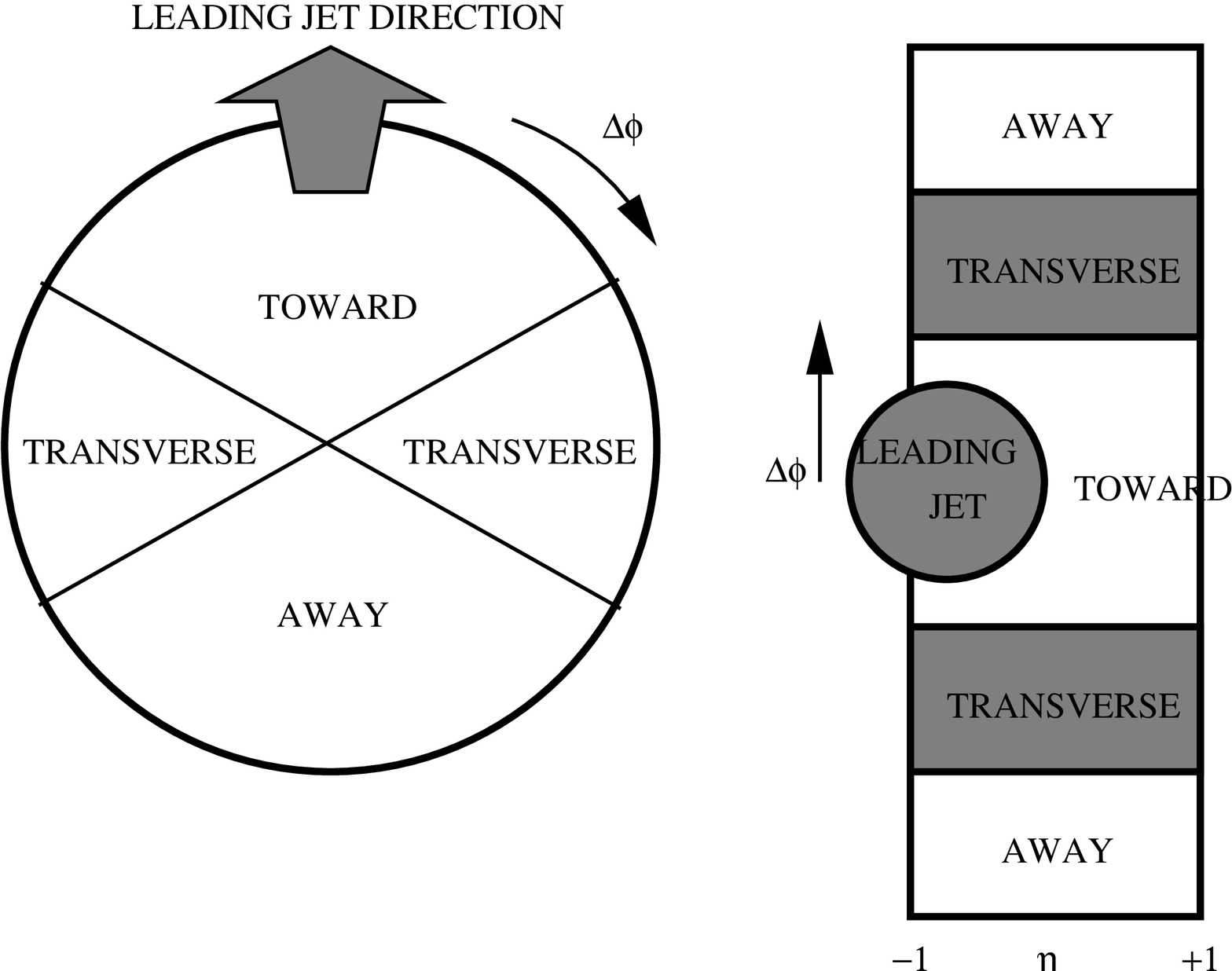,width=10cm}{`Toward', `away' and `transverse' 
regions defined with respect to the azimuthal direction of the leading 
charged jet.\label{fig_regions}}

  As shown in fig.~\ref{fig_regions}, three areas in the $\eta-\phi$ space
are defined with respect to the azimuthal direction of the leading charged
jet. $\eta$ is restricted to be between $-1$ and $+1$, and the three
regions are defined according to the azimuthal angle. The `toward'
direction is centred around the leading jet direction and is within
$\pi/3$ radians from the jet direction. The `transverse' direction is
between $\pi/3$ and $2\pi/3$ radians away from the jet direction. The
`away' direction is more than $2\pi/3$ radians away from the jet
direction. The three regions have equal area of $4\pi/3$ in the
$\eta-\phi$ space.

  The quantities considered in ref.~\cite{chargedjet}, that characterize
the underlying hadronic activity, are the average scalar sum
$p_T^\mathrm{sum}$ of $p_T$ of charged particles with $p_T$ greater than
0.5 GeV, and the charged particle multiplicity $N_\mathrm{chrg}$, in the
three regions described above. These quantities were plotted against the
$p_T$ of the leading charged jet.

  Out of the three regions, the region that is most sensitive to the
underlying event is the transverse direction.

 \subsection{Current status of HERWIG based simulation}

 \FIGURE[ht]{\epsfig{file=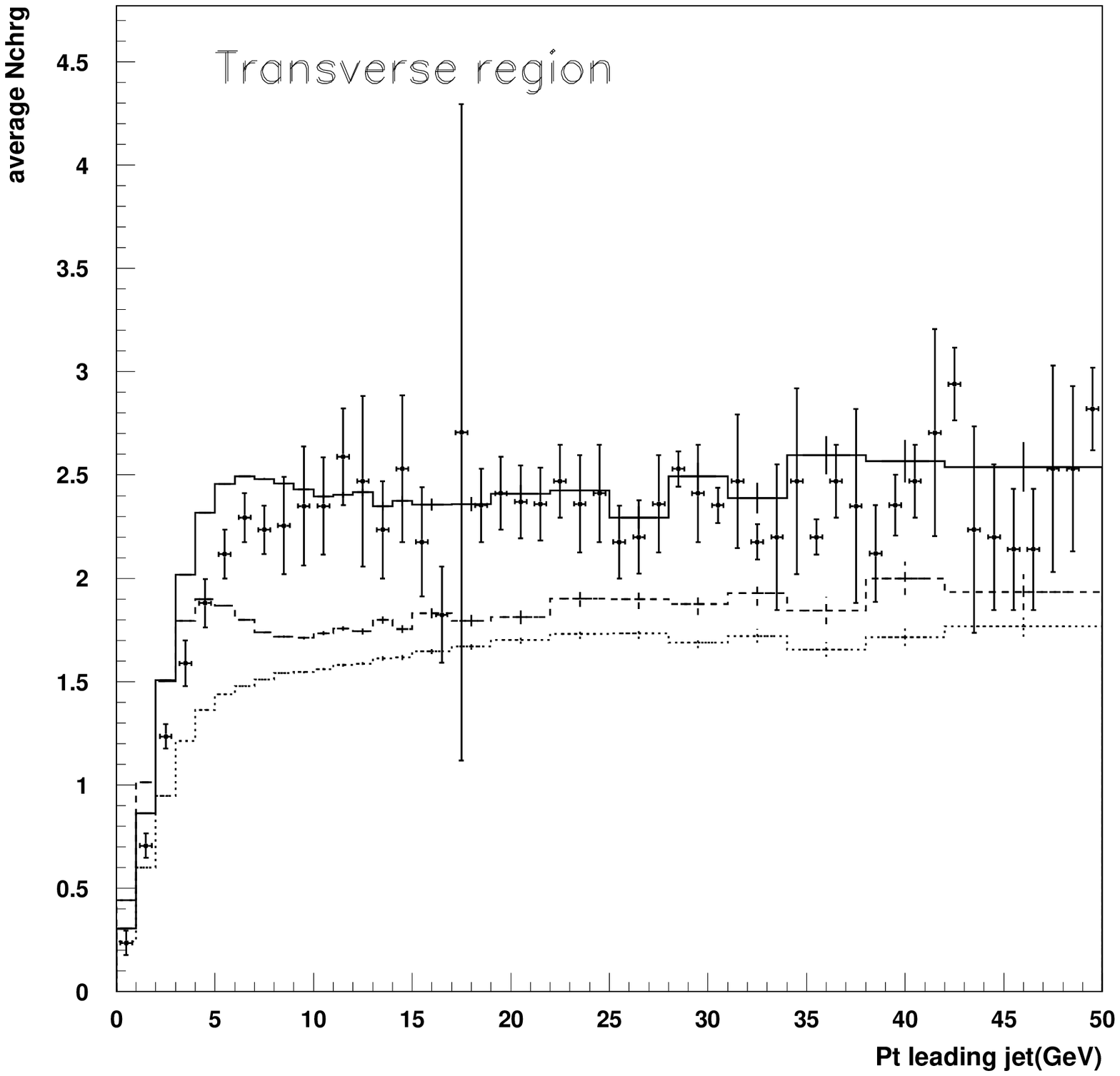,width=7.45cm}
             \epsfig{file=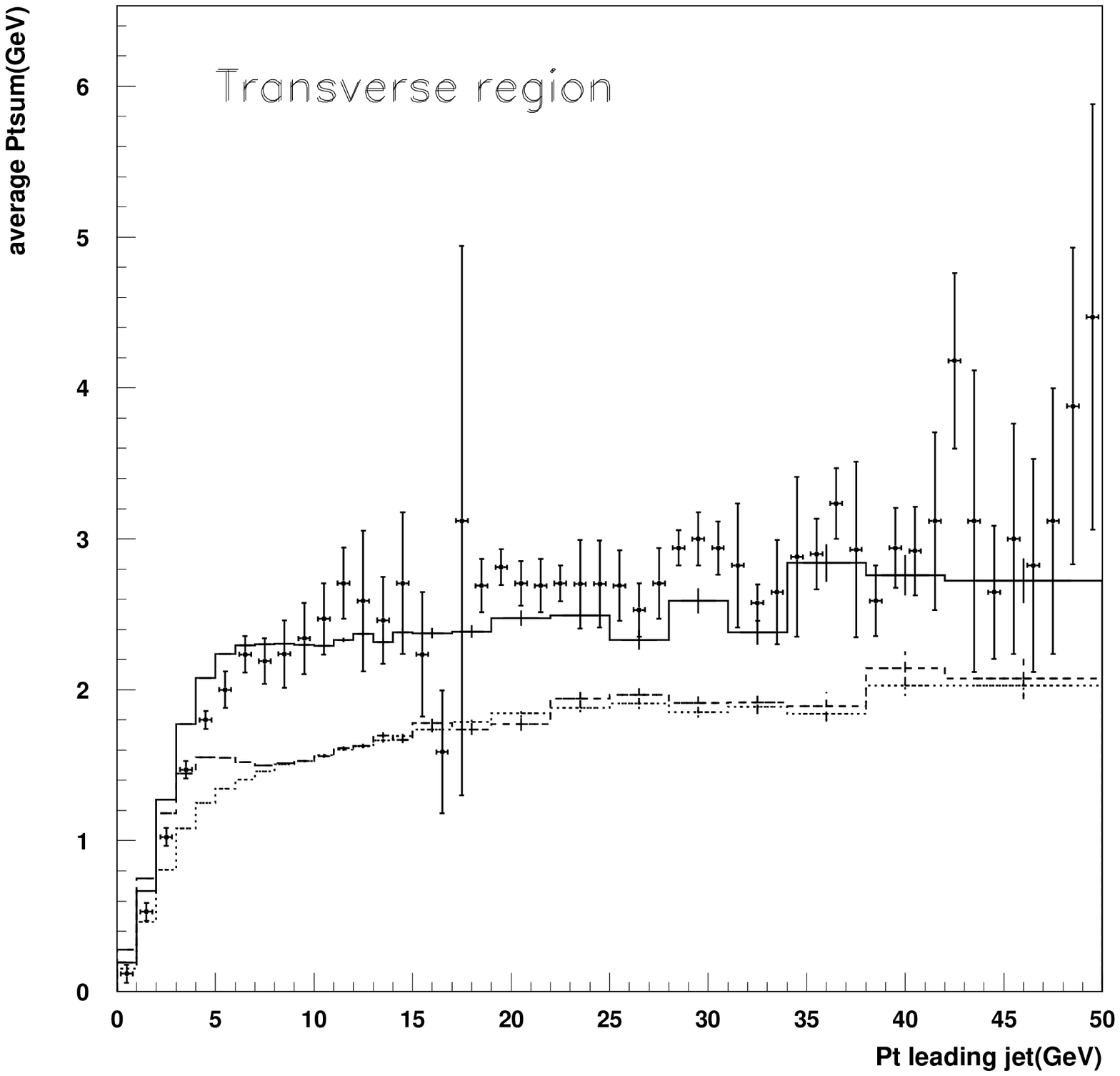,width=7.45cm}
 \caption{Comparison of `transverse' hadronic activity, in terms of
$N_\mathrm{chrg}$ (left) and $p_T^\mathrm{sum}$ (right), in various
HERWIG-based simulations, taken from ref.~\cite{borozan}. HERWIG+`Eikonal'
model (solid line), HERWIG+`Underlying Event' model (solid dashed),
HERWIG+`Multiparton Hard' model (dotted). Experimental data is shown with
error-bars.\label{fig_eikactivity}}}

  Fig.~\ref{fig_eikactivity}, taken\footnote{We thank the authors of
ref.~\cite{borozan} for their kind permission to reproduce their figure
here.} from ref.~\cite{borozan}, shows the current status of the fits to
the Tevatron data using HERWIG-based simulations.

  The default HERWIG curve is very much below the experimental numbers,
but this is subject to the tuning of parameters. The numbers shown in
ref.~\cite{chargedjet} are in better agreement, although still being lower
than the experimental result.

  The `multi-parton hard' model is the result of simulation using JIMMY
\cite{jimmy}. The simulation is based on the overlap-function approach.  
The plot shown has $p_T^\mathrm{min}=3$ GeV. The level of hadronic
activity is not sufficiently high, and better agreement with data is
obtained when the inclusive cross section is boosted by lowering
$p_T^\mathrm{min}$ to 2 GeV.

  The `eikonal' model is an extension of the JIMMY approach where the 
$gg\to gg$ scatters below $p_T^\mathrm{min}=3$ GeV are also generated by 
modelling the inclusive cross section as an exponential function of 
$p_T^2$. The gluon structure function is assumed to have a $\propto1/x$ 
behaviour. The dependence on $p_T^\mathrm{min}$ is, according to their 
claim \cite{borozan}, weakened, but is nevertheless present. The numbers 
drop considerably when $p_T^\mathrm{min}$ is lowered to 2 GeV. In our 
opinion, this is simply because the inclusive cross section in reality 
behaves more as a power, $\sim p_T^{-3}$, than an exponential.

  The `eikonal' model fits the data best out of the three approaches. This
is because there is not sufficient hadronic activity in the other two
approaches and this model adds a large amount of `soft' activity from
below $p_T^\mathrm{min}$.

  As far as the shape of the fit is concerned, there are two remaining
problems that are not rectified, which are:
 \begin{enumerate}
  \item When $p_T^\mathrm{min}$ is adjusted to fit $N_\mathrm{chrg}$,
there is insufficient $p_T^\mathrm{sum}$.
  In other words, the charged particles in the underlying event carry too
little $p_T$.
  \item The slope at low $p_T$ is too steep, i.e., there is too much
activity when $p_T$ of the leading jet is small. In other words, the
underlying event is too uniform.
 \end{enumerate}

  The above two points seem to indicate two things.

  First, the description of fragmentation in low $p_T$ scatters is not
adequate.

  Second, the excessive uniformness of the simulated underlying event
suggests that the idea of employing the cross section below
$p_T^\mathrm{min}$ might actually not reflect the true nature of the
underlying event.
  This same point applies to the model intrinsic to HERWIG, which also
seems to have too much activity implicitly assigned to low $p_T$ physics.

  There is a finding in ref.~\cite{sjostvanzijl} that, in the PYTHIA
framework, one needs a double-Gaussian overlap function with a `hard core'
in order to fit the data.
  We note that this is possibly another phenomenological indication of the
non-uniformity of the observed underlying event.

  One possibility for large $p_T^\mathrm{min}$ is that it should be kept
high at the scale of the Regge-perturbative `transition', or where colour
singlet interactions become more dominant.
  This would seem to be in agreement with the observation made earlier
that $\sigma_\mathrm{eff}$ is much smaller in reality than is supposed in
ref.~\cite{borozan}. If $\sigma_\mathrm{eff}$ is smaller, larger activity
can be obtained for smaller $\sigma^\mathrm{inc}$.

 \EPSFIGURE[ht]{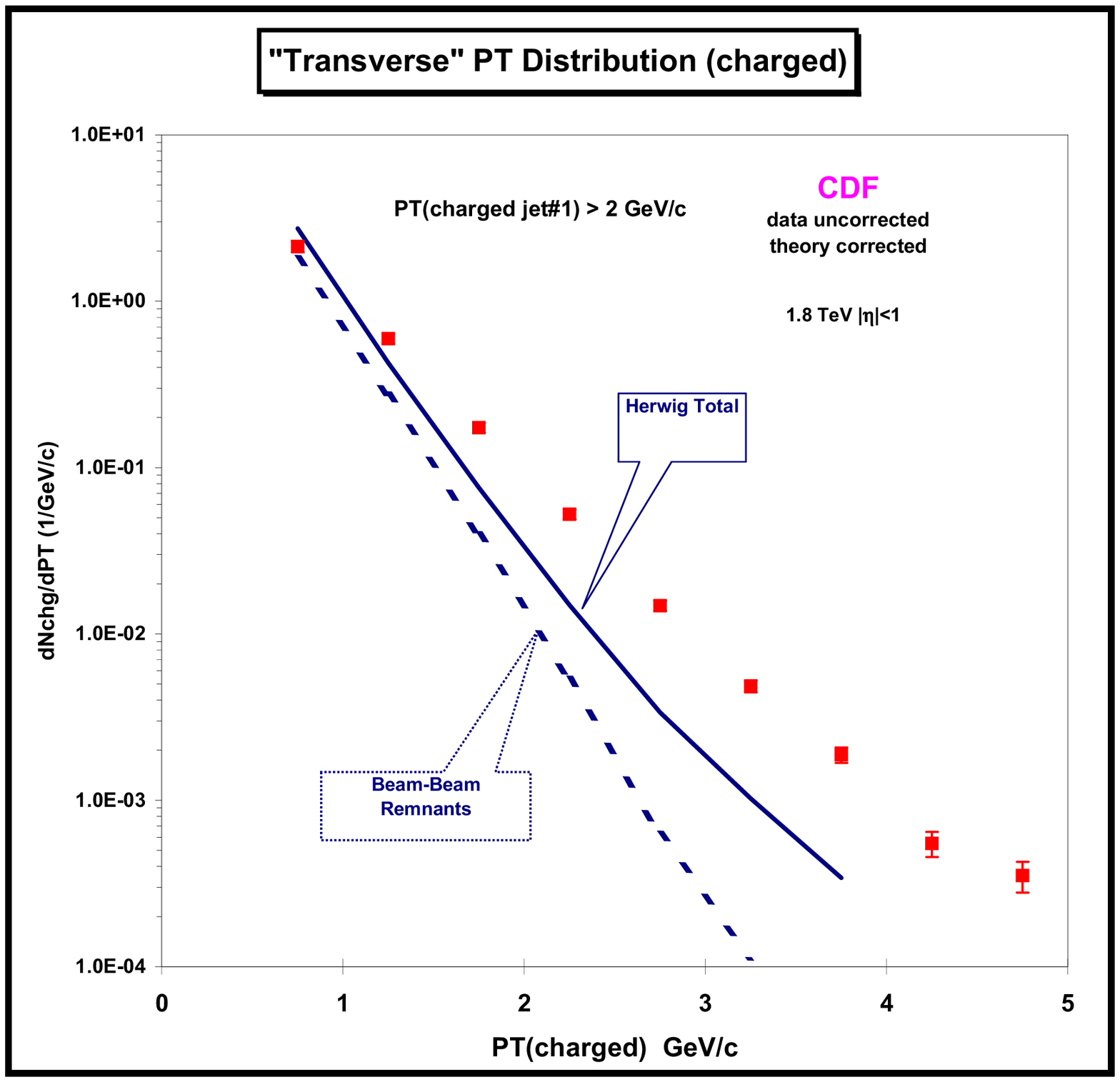,width=12cm}
 {Transverse momentum distribution of charged particles ($p_T>0.5$
GeV,$|\eta|<1$) in the `transverse' region, for leading jet $p_T>2$ GeV.  
Experimental data is compared against the HERWIG prediction. The dashed
curve shows the underlying event contribution. Figure taken from
ref.~\cite{chargedjet}.\label{fig_cdfpttracks}}

  In ref.~\cite{chargedjet}, in addition to the measurement of
$N_\mathrm{chrg}$ and $p_T^\mathrm{sum}$, there is also the measurement of
the $p_T$ of individual charged tracks in the `transverse' region. This is
reproduced\footnote{We thank the authors of ref.~\cite{chargedjet} for
their kind permission to reproduce their figure here.} in
fig.~\ref{fig_cdfpttracks} and we see that the model intrinsic in HERWIG
produces too many soft tracks.
  In ref.~\cite{chargedjet}, a study was made using the other generators,
and the result was that PYTHIA gives slightly better description of the
$p_T$ distribution although the slope is still too steep.

  One can easily confirm that the excessive softness of the charged tracks
shown in fig.~\ref{fig_cdfpttracks} is quantitatively consistent with our
earlier observation that when $p_T^\mathrm{min}$ is adjusted to fit
$N_\mathrm{chrg}$, there is insufficient $p_T^\mathrm{sum}$.

 \subsection{`Perturbative' nature of the underlying event}

  We remarked above that one of the reasons for the discrepancies between
theory and experiment may be that the scatters take place at larger $p_T$
than is usually supposed, with $p_T^\mathrm{min}$ being determined by the
Regge characteristic scale. If so, one would expect that, to a good
approximation, each scatter may be calculated using perturbation theory.

  To justify this claim in terms of the underlying theory, one needs to 
argue that the scatters from the Regge region do not contribute to the 
underlying event.
  This may be done in two steps by first showing that the Regge inclusive
cross section may be suppressed and then arguing that the fragmentation in
this region contributes less hadronic activity.

  In the perturbative region, for $p_T^\mathrm{min}$ reasonably small,
$\mathcal{O}(>2)$ GeV, the dependence of the inclusive cross section on
$p_T^\mathrm{min}$ is found to be roughly
$\sigma^\mathrm{inc}\sim(1/p_T^\mathrm{min})^{\sim3}$. We have found
numerically that the hadronic activity in default HERWIG also goes as
$(1/p_T^\mathrm{min})^{2\sim3}$.

  On the other hand, if the low $p_T$ scattering cross section is
determined by Regge dynamics, one may expect a form of the differential
cross section that goes as $d\sigma^\mathrm{inc}/dp_T^2\sim
p_T^{2\epsilon}$, with $\epsilon$ being a suitable Pomeron intercept.
  Hence the low $p_T$ cross section is suppressed compared with the
perturbative behaviour.
  We do not yet specify the nature of this Pomeron here, but our analysis
is inspired by the hard Pomeron picture of
refs.~\cite{donnachielandshoff,dlhardpomeron}.

 \EPSFIGURE[ht]{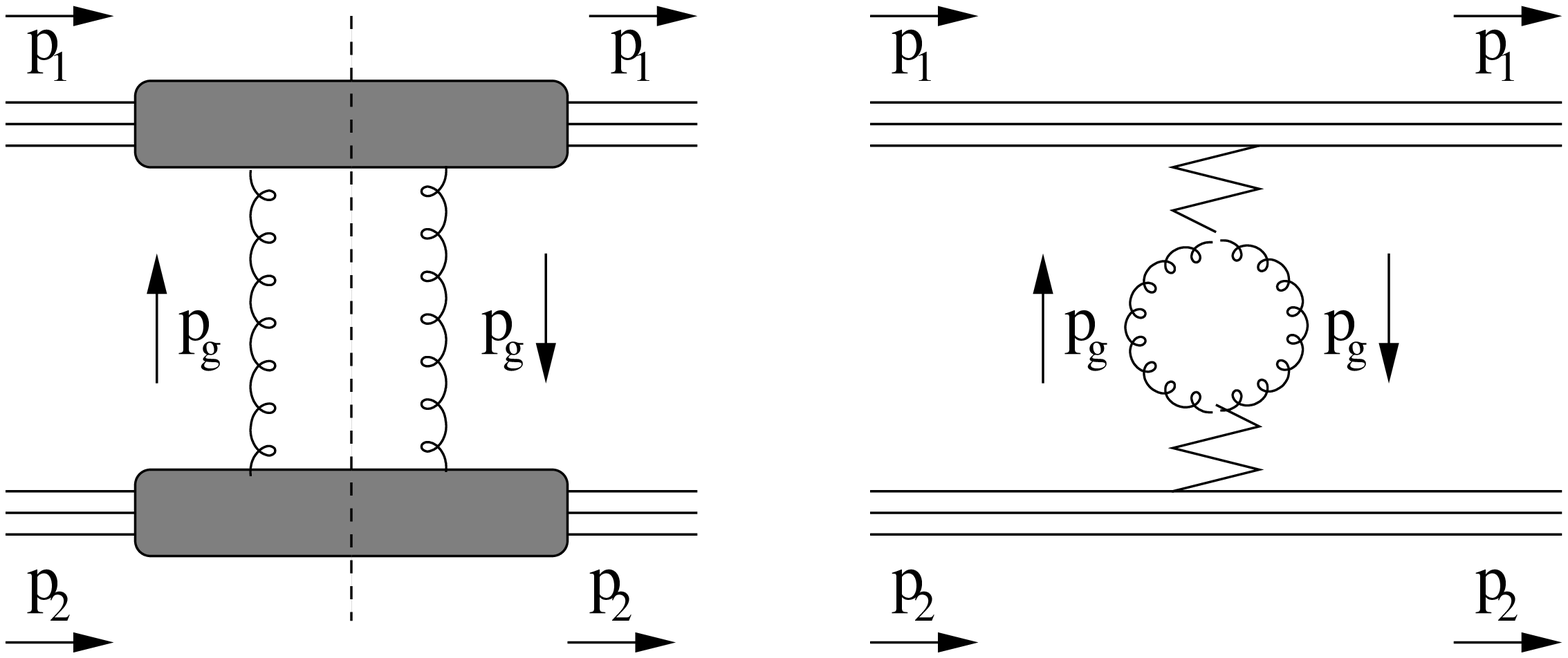,width=14cm}{The low $p_T$ colour-octet exchange
cross section (left) can be represented as the imaginary part of the
bubble insertion in the Pomeron exchange diagram
(right).\label{fig_bubble}}

  One possible picture which produces this $p_T$ dependence is illustrated
in fig.~\ref{fig_bubble}. We represent the `resolved' cross section by the
colour-octet gluon exchange cross section, and this is written as the
imaginary part of the bubble insertion in the Pomeron exchange diagram.

  We write the helicity and colour conserving gluon--proton amplitude at
zero momentum transfer as:
 \begin{equation}
  \mathcal{A}(g^A_{\lambda_g}p_{\lambda_p}\to
  g^A_{\lambda_g}p_{\lambda_p})\Big|_{t=0}=i\beta_{gp}
  \left(\frac{M_{pg}^2}{s_0}\right)^{1+\epsilon}.
 \end{equation}
  We have defined $\beta_{gp}$ as a dimensionless quantity. It is
proportional to the product of the dimension--$(-1)$ Pomeron couplings to
the gluon and to the proton. $\epsilon$ is the Pomeron intercept.
  $s_0$ is the Regge characteristic scale and is about 4 GeV$^2$
\cite{donnachielandshoff}. As this is an overall factor only, the
normalization may as well be traded between $\beta_{gp}$ and $s_0$.

  In ref.~\cite{dlpomform}, it is argued that the Pomeron coupling to
off-shell partons cannot be considered to be point-like and that one
should introduce a form factor with scale $\mathcal{O}(1$~GeV).
  Thus the above formulation is valid only below this scale. Above this
scale, the form factor should suppress the cross section in such a way to
reproduce the perturbative result
$\sigma^\mathrm{inc}\sim(1/p_T^\mathrm{min})^{\sim3}$. This picture is in
accord with the recent studies \cite{dlhardpomeron}.

  The whole proton--proton amplitude at zero momentum transfer then
follows:
 \begin{equation}
  \mathcal{A}_\mathrm{bubble}(p_{\lambda_1}p_{\lambda_2}\to
  p_{\lambda_1}p_{\lambda_2})\Big|_{t=0}=2i\beta_{gp}^2(N_C^2-1)
  \int\frac{d^4p_g}{(2\pi)^4p_g^4}
  \left(\frac{M_{1g}^2}{s_0}\right)^{1+\epsilon}
  \left(\frac{M_{2g}^2}{s_0}\right)^{1+\epsilon}.
  \label{eqn_pptopp}
 \end{equation}
  The factor 2 in the front is for the two transverse polarizations of the
gluon. The Pomeron coupling is thought to conserve helicity.
  We have added the subscript to $\mathcal{A}_\mathrm{bubble}$ to
emphasize the point that we are not calculating the total elastic and
inelastic cross section but only the `resolved' cross section, here being
identified with the colour octet transfer contribution.
  The $pp\to X$ cross section is then calculated by the optical theorem:
 \begin{equation}
  \sigma_\mathrm{res}^\mathrm{inc}(pp\to X)=
  \frac1{s_\mathrm{CM}}\mathrm{Im}\mathcal{A}_\mathrm{bubble}(pp\to pp)
  \Big|_{t=0}.
 \end{equation}

  Neglecting the proton mass, let us define the four-momenta as follows:
 \begin{eqnarray}
  p_1&=&\frac{\sqrt{s_\mathrm{CM}}}2\left(1,0,0,1\right),  \nonumber\\
  p_2&=&\frac{\sqrt{s_\mathrm{CM}}}2\left(1,0,0,-1\right), \nonumber\\
  p_g&=&\left(E,\mathbf{p}_T,-p_z\right).
 \end{eqnarray}
  $p_z$ is positive definite and satisfies $p_z>|E|$.

  We express the phase space in terms of $\mathbf{p}_T$ and rapidity as:
 \begin{equation}
  d^4p_g=\frac12d^2\mathbf{p}_Tdt_gdy=\frac\pi2d\mathbf{p}_T^2dt_gdy.
 \end{equation}
  Here $t_g=p_g^2$. We added the subscript to avoid confusion with the 
overall momentum transfer in the amplitude, which is zero.
  We define rapidity as:
 \begin{equation}
  y=\frac12\ln\frac{p_z+E}{p_z-E}.
 \end{equation}
  The two invariant masses squared in eqn.~(\ref{eqn_pptopp}) are given
by:
 \begin{eqnarray}
  M_{1g}^2 &=& p_g^2+\sqrt{s_\mathrm{CM}}(p_z+E)
   = t_g+e^y   \sqrt{s_\mathrm{CM}(-t_g-\mathbf{p}_T^2)},\nonumber\\
  M_{2g}^2 &=& p_g^2+\sqrt{s_\mathrm{CM}}(p_z-E)
   = t_g+e^{-y}\sqrt{s_\mathrm{CM}(-t_g-\mathbf{p}_T^2)}.
  \label{eqn_invmasses}
 \end{eqnarray}
  As $M_{1g}^2,M_{2g}^2>0$, we have the inequalities:
 \begin{eqnarray}
  |y| &<& \ln\left(\frac{\sqrt{s_\mathrm{CM}(-t_g-\mathbf{p}_T^2)}}
	          {-t_g}\right), \nonumber\\
  0<\mathbf{p}_T^2&<&-t_g-t_g^2/s_\mathrm{CM}.
  \label{eqn_yptlimits}
 \end{eqnarray}

  The gluon virtuality terms $t_g$ in eqn.~(\ref{eqn_invmasses}) are small
and so let us neglect them. This is admissible as the cross section is
small-$x$ dominated in the structure function language.
  The $y$ integration yields a logarithm which is approximately
$\ln(s_\mathrm{CM}/s_0)$. The remaining integration in
eqn.~(\ref{eqn_pptopp}), expressed in terms of
$(d\sigma^\mathrm{inc}_\mathrm{res}/dt_g)$, is merely:
 \begin{equation}
  \frac{d\sigma^\mathrm{inc}_\mathrm{res}}{dt_g}\approx
  \frac{\beta_{gp}^2(N_C^2-1)}{16\pi^3s_0t_g^2}
  \left(\frac{s_\mathrm{CM}}{s_0}\right)^\epsilon
  \ln\left(\frac{s_\mathrm{CM}}{s_0}\right)
  \int d|m_T^2|\left(\frac{|m_T^2|}{s_0}\right)^{1+\epsilon}.
 \end{equation}
  The proton helicities have been averaged over.
  We have defined $m_T^2=t_g+\mathbf{p}_T^2$ which is negative definite
from eqn.~(\ref{eqn_yptlimits}). The $m_T^2$ integration is of course
trivial. Neglecting the small contribution from the lower limit of
integration, we obtain:
 \begin{equation}
  \frac{d\sigma^\mathrm{inc}_\mathrm{res}}{dt_g}\approx
  \frac{\beta_{gp}^2(N_C^2-1)}{16\pi^3s_0^2(2+\epsilon)}
  \left(\frac{s_\mathrm{CM}}{s_0}\right)^\epsilon
  \left(\frac{-t_g}{s_0}\right)^{\epsilon}
  \ln\left(\frac{s_\mathrm{CM}}{s_0}\right).
  \label{eqn_dsigdtfinal}
 \end{equation}
  Thus the cross section has a $(-t_g)^\epsilon$ behaviour and falls to
zero at zero gluon virtuality rather than increasing indefinitely.
  $(-t_g)$ can be loosely exchanged with $\mathbf{p}_T^2$ and hence we 
have demonstrated the power behaviour of the low $p_T$ cross section 
claimed earlier.

  It is possible to interpret the above discussion in terms of the hard
Pomeron picture of ref.~\cite{dlhardpomeron}, with $\epsilon\approx0.4$.  
This provides a means of fixing the normalization of
eqn.~(\ref{eqn_dsigdtfinal}), by using the gluon structure function.

  Let us consider the first diagram of fig.~\ref{fig_bubble} but with the
upper proton replaced by the gluon. We write the $gg\to gg$ amplitude at
zero momentum transfer, averaged over the first gluon colour and helicity,
as:
 \begin{equation}
  \mathcal{A}(g g^A_{\lambda_g}\to g g^A_{\lambda_g}) \Big|_{t=0}
  =iM^2_{gg}\widehat{\sigma}_{gg}(M^2_{gg}).
 \end{equation}
  The whole amplitude is then:
 \begin{eqnarray}
  \mathcal{A}(gp_{\lambda_p}\to gp_{\lambda_p})\Big|_{t=0}
  &=&2i\beta_{gp}(N_C^2-1)
  \int\frac{d^4p_g}{(2\pi)^4p_g^4}
  M_{1g}^2\widehat{\sigma}_{gg}(M^2_{1g})
  \left(\frac{M_{2g}^2}{s_0}\right)^{1+\epsilon}\nonumber\\
  &=&i\frac{\beta_{gp}(N_C^2-1)}{16\pi^3}
  \int dt_gd\mathbf{p}_T^2dy M_{1g}^2\widehat{\sigma}_{gg}(M^2_{1g})
  \left(\frac{M_{2g}^2}{s_0}\right)^{1+\epsilon}.
 \end{eqnarray}
  We are interested in the structure function
$f_g(x,Q^2)=(\partial\sigma/\partial x)/\widehat\sigma$. We define
$x=(p_Z+E)/\sqrt{s_\mathrm{CM}}$ and omit $t_g$ in 
eqn.~(\ref{eqn_invmasses}) as before, to obtain:
 \begin{equation}
  f_g(x,Q^2)\approx\frac{\beta_{gp}(N_C^2-1)}{16\pi^3}
  \int_{t_g=0}^{t_g=Q^2} \frac{dt_g}{t_g^2}d|m_T^2|
  \left(\frac{-m_T^2}{s_0x}\right)^{1+\epsilon}.
 \end{equation}
  Hence, neglecting the lower limit of $m_T^2$ integration as before, we
obtain:
 \begin{equation}
  f_g(x,Q^2) \approx
  \frac{\beta_{gp}(N_C^2-1)}{16\pi^3(1+\epsilon)(2+\epsilon)}
  \left(\frac{Q^2}{s_0x}\right)^{1+\epsilon}.
  \label{eqn_strfn}
 \end{equation}
  We may compare this with the form given in ref.~\cite{dlhardpomeron}:
 \begin{equation}
  xg(x,Q^2)\sim0.95(Q^2)^{1+\epsilon_0}
  (1+Q^2/0.5)^{-1-\epsilon_0/2}x^{-\epsilon_0}.
  \label{eqn_strfnpvl}
 \end{equation}
  The units for $Q^2$ are in GeV$^2$.
  This form is claimed to be valid in the range $5<Q^2<500$ GeV$^2$.
  $\epsilon_0$ is the hard Pomeron intercept which, in our notation, is
$\epsilon\approx0.4$.
  Because of the presence of the extra denominator factor
$(1+Q^2/0.5)^{-1-\epsilon_0/2}$, the normalization is ambiguous. For
instance, if we extend their claimed range of applicability to the limit
of very low $Q^2$ and match the two expressions at $Q^2=0$, we obtain a
cross section which is unreasonably large.
  Let us match the two equations at the lowest end, $Q^2=5$ GeV$^2$, of
the range of validity of eqn.~(\ref{eqn_strfnpvl}). We obtain:
 \begin{equation}
  \frac{\beta_{gp}(N_C^2-1)}{16\pi^3(1+\epsilon)(2+\epsilon)}
  \sim0.05
 \end{equation}
  At $\sqrt{s}_\mathrm{CM}=1.8$ TeV, we then obtain from
eqn.~(\ref{eqn_dsigdtfinal}):
 \begin{equation}
  \frac{d\sigma^\mathrm{inc}_\mathrm{res}}{dt_g}\approx
  30\ \mathrm{mb\ GeV}^{-2}
  \left(\frac{-t_g}{s_0}\right)^{\epsilon}.
  \label{eqn_dsigdtnum}
 \end{equation}
  As stated above, there is ambiguity in the normalization due to our lack
of knowledge about the structure function at very low $Q^2$.

 \subsection{Fragmentation in low $p_T$ scatters}

  In addition to the Regge suppression of low $p_T$ scattering rate, we 
note that fragmentation may also be affected by Regge dynamics.

  One characteristic behaviour is Reggeization. Partons that are emitted
in the perturbative phase are coloured objects with colour partners, but
in the Regge phase, they may become, in some sense, colour singlet.
  If this is the case, large rapidity gaps would be formed per scatter and
hence the contribution to the underlying event would be reduced.

  This effect is difficult to quantify, but since the leading effect is to 
suppress the low $p_T$ contribution, this provides another justification 
for neglecting it.

  In ref.~\cite{odagirihadronization}, we proposed that the splitting of 
large clusters in the cluster-hadronization model 
\cite{cluster_hadronization} of HERWIG can be connected to a modified QCD 
coupling that governs those emissions which are considered unresolved in 
the parton-shower phase.
  `Clusters' are colour-singlet quark-antiquark units which result from
the parton shower process followed by a forced $g\to q\bar q$ splitting.
In HERWIG, they subsequently decay isotropically into hadrons according to
the phase space weight.

  The effect of this modification is not particularly large for processes
occuring at large hard-process scale, but it is potentially significant
for soft processes such as the underlying event.
  Therefore the study of the underlying event, in particular relatively
inclusive quantities such as the $p_T$ distribution of charged particles
shown in fig.~\ref{fig_cdfpttracks}, could provide a testing ground for
models of hadronization.

  One possibility that follows from the modified picture of hadronization 
is that because of the reduced phase space for parton emission in soft 
scatters, the resulting cluster size is typically larger than in the high 
$p_T$ processes. This is discussed in the next section, together with the 
possibility that this leads to different individual hadron yields.

 \section{Simulation and discussions}\label{sec_simulation}

  Fitting the experimental data with the simulation tools involves, in
general, not a small number of tunable parameters. However, they are
reducible to three components. First, the `effective cross section'
$\sigma_\mathrm{eff}$. Second, the inclusive cross section controlled by
$p_T^\mathrm{min}$. Third, the distribution of the number of scatters
controlled by the overlap function. Normally $\sigma_\mathrm{eff}$ is not
an explicitly tunable parameter, so that one adjusts $p_T^\mathrm{min}$ to
adjust the amount of hadronic activity. The distribution of the number of
scatters then affects how much the multiple-scattering contribution
fluctuates. In addition, one sometimes makes the cut-off at
$p_T^\mathrm{min}$ smooth, and this also affects the shape of
distributions.

  We now abandon the overlap function and adopt a naive Poisson
distribution. We hence have only one parameter, $p_T^\mathrm{min}$, that
can be tuned to fit the amount of hadronic activity, and have an
environment to study the fragmentation scheme dependence.

  Although $p_T^\mathrm{min}$ is a tunable parameter, we should not set it
too far from the onset of Regge physics. A reasonable value may be in the
range $1-5$ GeV and so let us adopt 3 GeV for now.
  For $\sigma_\mathrm{eff}$, we adopt the measured value of 14.5 mb.
  We carry out the simulation simply by generating a number of QCD
$(2\to2)$ scatters on top of one another, without considering the
constraint from the conservation of energy.

  The experimental data is uncorrected and the theoretical numbers are
corrected by removing, on average, 8\% of the charged tracks.

 \FIGURE[ht]{
 \centerline{\epsfig{file=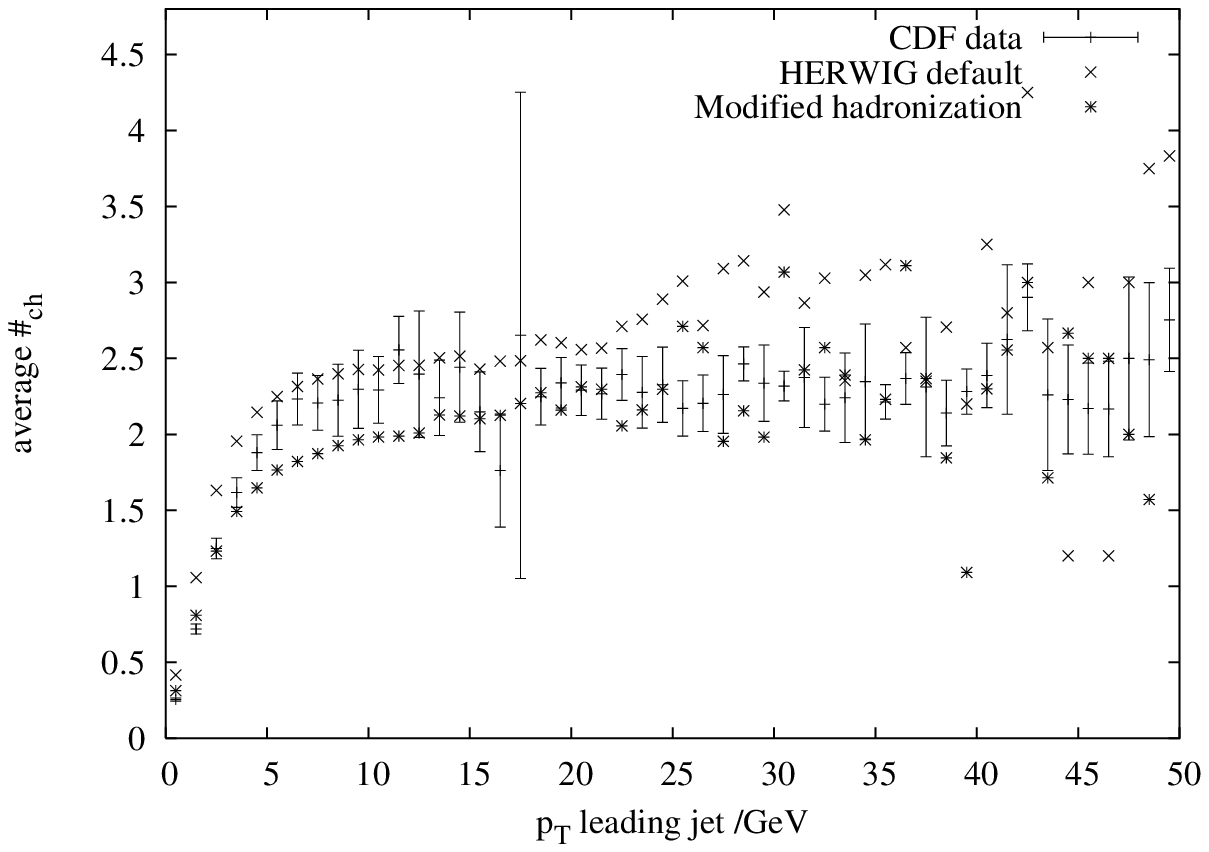,width=11cm}}
 \centerline{\epsfig{file=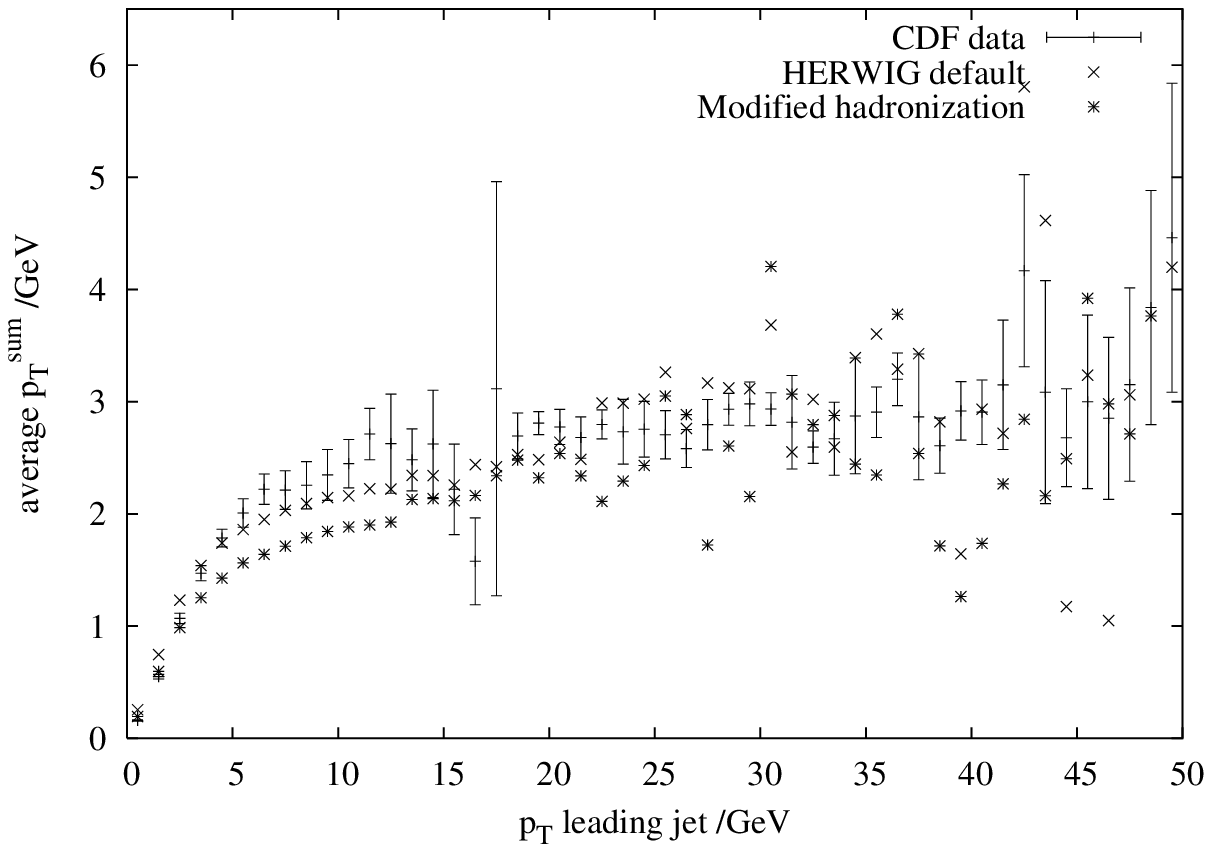,width=10.8cm}}
 \caption{Comparison of `transverse' hadronic activity, in terms of
$N_\mathrm{chrg}$ (above) and $p_T^\mathrm{sum}$ (below). Simulation at
$p_T^\mathrm{min}=3$ GeV and $\sigma_\mathrm{eff}=14.5$ mb. The two
hadronization schemes are explained in the text. Experimental data is
shown with error-bars. Simulation sample size is one million events.
Error-bars are not shown.
 \label{fig_simpleresult}}}

  In fig.~\ref{fig_simpleresult}, we show the result of a simple
simulation using HERWIG. The inclusive cross section is found to be about
27 mb so that about two scatters take place per event on average. With
this value of $p_T^\mathrm{min}$, the normalization is found to match that
of experimental data. We note that in ref.~\cite{borozan},
$\sigma_\mathrm{eff}$ is about 60 mb but the inclusive cross section is
larger and hence the average number of scatters per event is 2.4, not too
dissimilar to the case studied here.

  Two hadronization schemes are used. The first is the default cluster
hadronization scheme of HERWIG. The second is the scheme proposed in
ref.~\cite{odagirihadronization} which uses a Gaussian approximation for
low energy $\alpha_S$ to describe the splitting of large clusters.

  The normalization is smaller for the modified cluster splitting
procedure based on the low-energy Gaussian $\alpha_S$. This is because
when very large clusters are split, the modified procedure results in
lighter daughter clusters and hence smaller multiplicity. We note that,
for the same reason, the $p_T^\mathrm{min}$ dependence of the underlying
hadronic activity is milder in this approach. This is welcome as it
reduces the theoretical uncertainty due to $p_T^\mathrm{min}$.

  The two discrepancies with experimental data mentioned in
sec.~\ref{sec_nature} are still present to some extent, though the slope
for low values of leading-jet $p_T$ is now in better agreement with
experimental data. The remaining problem is that the tracks are still too
soft.

 \EPSFIGURE[ht]{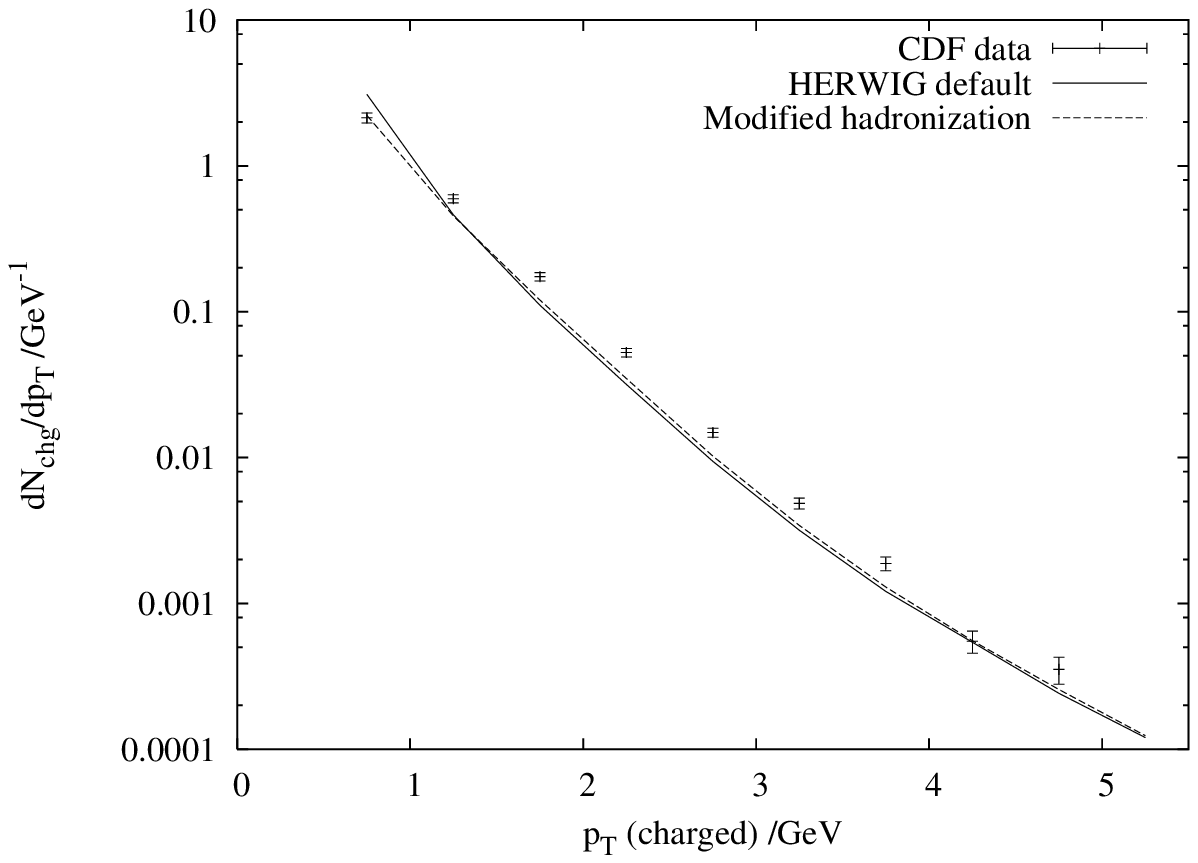,width=14cm}
 {Transverse momentum distribution of charged particles ($p_T>0.5$
GeV,$|\eta|<1$) in the `transverse' region, for leading jet $p_T>2$ GeV.  
Experimental data is compared against the HERWIG prediction for the two 
hadronization schemes.\label{fig_hwdefpttracks}}

  This last point is appreciated in a more quantitative fashion by
considering the $p_T$ distribution of charged tracks in the transverse
region, as shown in fig.~\ref{fig_hwdefpttracks}.

  There is better agreement between experimental data and simulation than
in fig.~\ref{fig_cdfpttracks}, but some discrepancy still remains. The
difference between the two hadronization schemes is small. The most
visible difference is for the point at $p_T$ between 0.5 and 1 GeV, where
the modified hadronization scheme results in fewer tracks, in better
agreement with experimental data.

  One possibility, as to the reason why the $p_T$ distribution of charged
tracks is too soft even with large $p_T^\mathrm{min}$, has to do with the
cluster splitting process.
  We proposed in ref.~\cite{odagirihadronization} that the splitting of
large clusters is due to the parton emission that is considered unresolved
in the usual parton-shower phase. If so, and considering that the scatters
take place mostly at very low $p_T$, the typical size of clusters from
scatters at low $p_T$ is expected to be larger due to the lack of phase
space for parton emission.

  Let us consider the cascade splitting of a cluster with mass
$M_\mathrm{init}$.
  From ref.~\cite{odagirihadronization}, for a cluster splitting process
$0\to12$ by unresolved parton emission at $p_T^2=Q^2$ we have:
 \begin{equation}
  Q^2=\frac{M_1^2M_2^2}{M_0^2},
 \end{equation}
  and the number of splittings is given by:
 \begin{equation}
  \left<\#_\mathrm{split}\right>\sim 
  I_0\log\frac{M_\mathrm{init}^2}{\left<Q^2\right>}.
  \label{eqn_splitmulti}
 \end{equation}
  Here $I_0$ is $C_F/\pi$ times the integral of low energy $\alpha_S$, and
is found to be about 0.5.
  Combining the above two equations and considering cascade decay with
final cluster masses $M_\mathrm{final}$, we obtain:
 \begin{equation}
  \left<\log\frac{M_\mathrm{final}^2}{Q^2}\right>
  = \frac1{I_0(1+1/\#_\mathrm{split})}.
  \label{eqn_meanclusmas}
 \end{equation}
  In high $p_T$ processes, clusters that are formed before the
cluster-splitting process have relatively low mass, with some large
clusters having masses extending up to $\mathcal{O}$(10 GeV). In this
case, the right hand side of eqn.~(\ref{eqn_meanclusmas}) is about 1.5 for
$\left<Q\right>\approx0.75$ GeV. Hence we obtain
$\left<M_\mathrm{final}\right>\approx1.5$ GeV. On the other hand, in the
large $M_\mathrm{init}$ limit, in other words
$\#_\mathrm{split}\to\infty$, we obtain $\left<M_\mathrm{final}\right>\to
2$ GeV.

  One simple way of modifying the typical cluster mass as a means to study
this aspect of hadronization dynamics is to change the HERWIG parameter
{\tt CLMAX}, which is the maximum allowed cluster mass.
  This is imposed even in the case of the modified algorithm.
  Clusters with mass greater than {\tt CLMAX} are split. We modify this,
somewhat arbitrarily, from the default value of 3.35 GeV to 5 GeV. We
adopt the low-energy $\alpha_S$ prescription for splitting clusters, as
otherwise there is no incentive for modifying {\tt CLMAX}.

 \EPSFIGURE[ht]{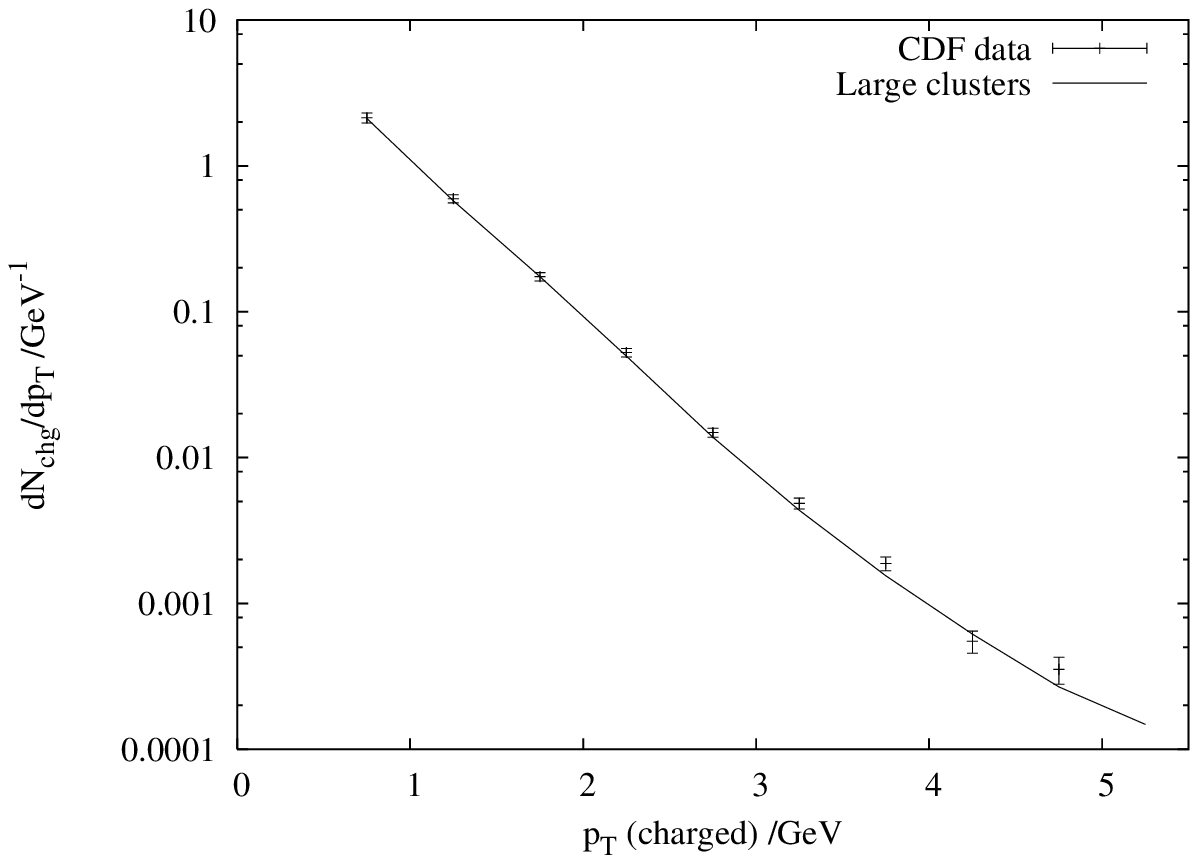,width=14cm}
 {Transverse momentum distribution of charged particles ($p_T>0.5$
GeV,$|\eta|<1$) in the `transverse' region, for leading jet $p_T>2$ GeV.  
Experimental data is compared against the HERWIG prediction for the
modified hadronization scheme with {\tt CLMAX} set to 5
GeV.\label{fig_hwlargecluspttracks}}

  The result, for the $p_T$ distribution of charged tracks, is shown in 
fig.~\ref{fig_hwlargecluspttracks}, and is in very good agreement with CDF 
data.

 \FIGURE[ht]{
 \centerline{\epsfig{file=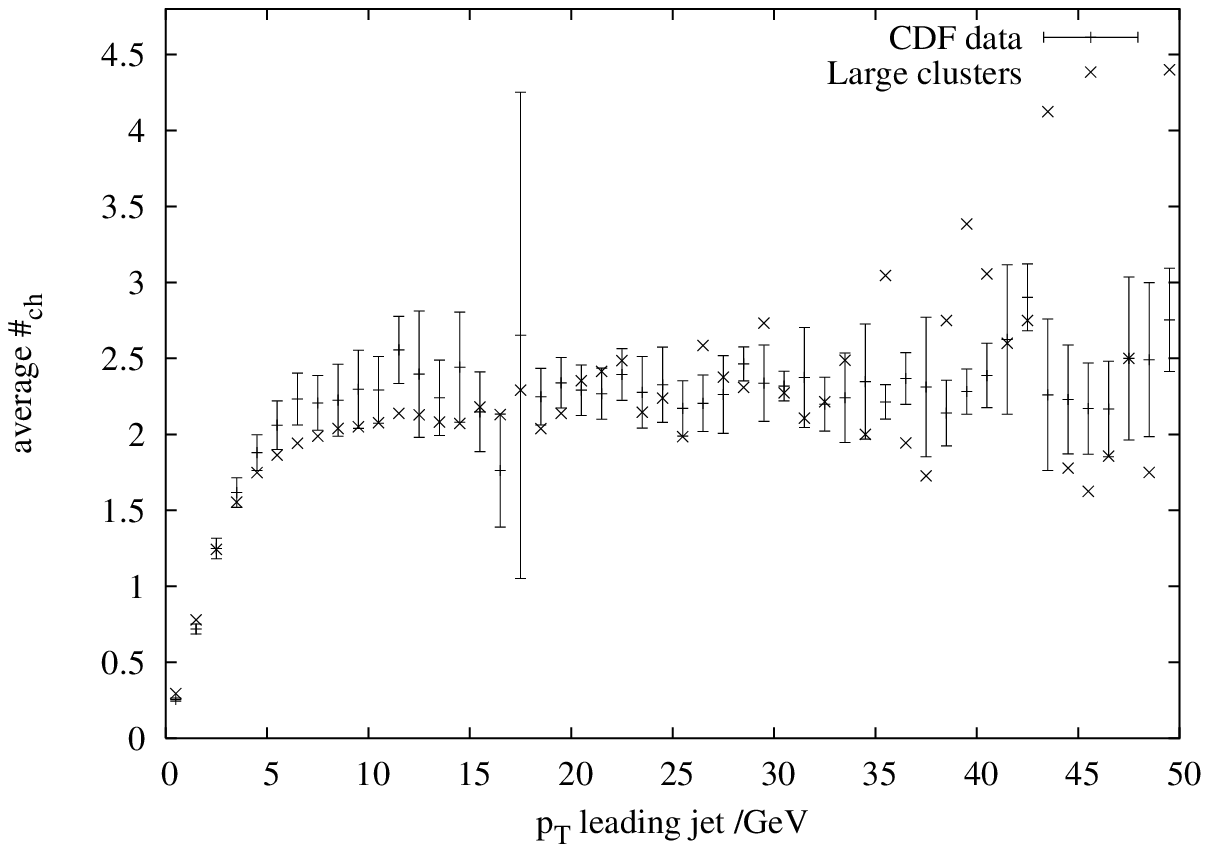,width=11cm}}
 \centerline{\epsfig{file=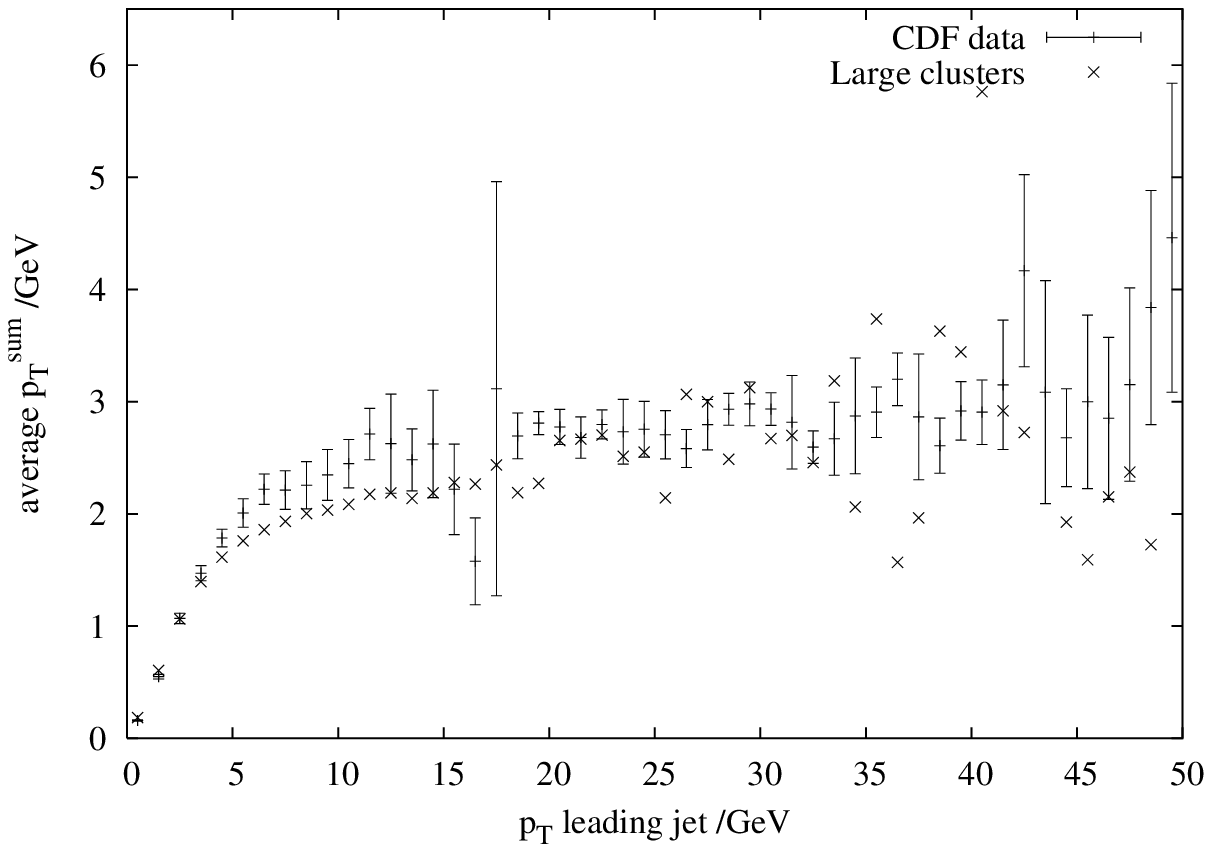,width=10.8cm}}
 \caption{Comparison of `transverse' hadronic activity, in terms of
$N_\mathrm{chrg}$ (above) and $p_T^\mathrm{sum}$ (below). Simulation at
with {\tt CLMAX} set to 5 GeV. Experimental data is shown with error-bars.
Simulation sample size is ten million events. Error-bars are not shown.
 \label{fig_largeclusresult}}}

  The leading-jet $p_T$ dependence of the transverse hadronic activity,
shown in fig.~\ref{fig_largeclusresult}, also shows good agreement with
data. In addition to the improved $p_T$ per charged track ratio, there is
also improvement to the shape of the low leading-jet $p_T$ region.

  The remaining difference may be due either to our incomplete description
of hadronization or to the inaccurate distribution of the number of
scatters.
  We have tested lowering $p_T^\mathrm{min}$ to 2.5 GeV and found that
this results in too many soft tracks in fig.~\ref{fig_hwlargecluspttracks}
so we believe that the wrong choice of $p_T^\mathrm{min}$ alone is not a
likely possibility to explain the imperfect fit.

  If the above speculations concerning the fragmentation in low $p_T$
scatters are correct, one would expect that the identified particle yields
may be different. With enhanced cluster mass, there is more phase space
for decay into heavier hadrons, for instance baryons, and hence we would
expect increased baryon--meson ratio. This could be confirmed by measuring
the proton--charged pion ratio $R_{p/\pi}$.

 \TABULAR[ht]{|c|c|c|c|}{
 \hline
 hadronization scheme 
 & $N_{p,\overline{p}}$ & $N_{\pi^\pm}$ & $R_{p/\pi}$ \\ \hline
 HERWIG default      & 3748 & 18056 & $0.208\pm0.004$ \\
 Gaussian $\alpha_S$ & 2698 & 15032 & $0.179\pm0.004$ \\
 Large clusters      & 4334 & 14474 & $0.299\pm0.005$ \\
 \hline
 }
 {The proton to charged pion ratio, for tracks with $|\eta|<1$ and 
$p_T>0.5$ GeV, in the three hadronization schemes adopted in this study. 
Sample size is 10000 scatters. $p_T^\mathrm{min}=3$ GeV. The errors quoted 
are due to the Monte Carlo statistics only.
 \label{tab_ptopi}}

  A simple simulation with a small statistics of 10000 scatters, sampling
particles with $p_T>0.5$ GeV and $|\eta|<1$ only, yields the numbers shown
in tab.~\ref{tab_ptopi}. We see that in the case with {\tt CLMAX} set to 5
GeV, there is about 50\% increase in the proton yield as compared to the
case with the default {\tt CLMAX} of 3.35 GeV.

  We note that even the unmodified case leads to substantially increased
yield for protons compared with yields in high $p_T$ collision. For
instance, the numbers from LEP at the $Z^0$ pole gives
$R_{p/\pi}\approx1/17$ \cite{pdgdata}.

  We should say that some caution is needed as regards the numbers, as the
identified particle yield for baryons is a poorly described quantity in
HERWIG. Hence although the above argument does indicate that the proton
yield is substantially enhanced in the underlying event, the exact numbers
for the yield should not be trusted.

  We propose the measurement of $R_{p/\pi}$, and other hadron yields, in
the `toward' and `transverse' regions as a function of the leading jet
$p_T$ as a way to understand the fragmentation properties of low $p_T$
scatters.

 \section{Conclusions and outlook}\label{sec_conclusions}

 \subsection*{Conclusions}

  We studied the underlying event in the multiple-scattering picture at
Tevatron.

  We started by noting some deficiencies of the current theoretical
description of the underlying event. We noted the possibility that the
scatters affecting the underlying event take place at above the
Regge-dominated regime and are therefore to a good approximation
calculable using perturbation theory. We also noted that the description
of fragmentation in low $p_T$ scatters needs to be improved.

  We carried out our simulation on HERWIG using a naive Poisson 
distribution for the number of scatters, choosing $p_T^\mathrm{min}=3$ GeV 
and $\sigma_\mathrm{eff}=14.5$ mb, and found that by modifying the 
hadronization algorithm as suggested in our previous study and by adopting 
a larger maximum cluster size parameter {\tt CLMAX}, a good agreement with 
data is obtained.

  If our picture is correct, one would expect an enhancement in the
baryon yield in the underlying event. We proposed that a measurement is
made of this effect by measuring the proton-to-pion ratio as a function of
the leading jet $p_T$. A confirmation of this effect, or otherwise, could
shed light both on the nature of the underlying event and on the nature of
the dynamics of hadronization.

 \subsection*{Outlook}

  A possible future improvement to our study may come from a better
understanding of the distribution of the number of scatters per event, or
in other words the mechanism of unitarization of scattering probabilities.
We desire further experimental study, in particular the measurement of
$\sigma_\mathrm{eff}$ on distinct experimental platforms, for instance in
resolved $\gamma p$ and $\gamma\gamma$ collisions, and at different
$\sqrt{s_\mathrm{CM}}$. The large $\sqrt{s_\mathrm{CM}}$ available at LHC
will give rise to larger mean number of scatters per event. This will be
significant aid to the study of the distribution of the number of
scatters.

  Additionally to the results presented herein, we have made a 
small-statistics calculation for the underlying event at LHC using the 
same $\sigma_\mathrm{eff}$ and $p_T^\mathrm{min}$. The flattening-off of 
the transverse hadronic activity occurs at larger trigger-jet $p_T$. For 
sufficiently large $p_T$, both $N_\mathrm{chrg}$ and $p_T^\mathrm{sum}$ 
are found to be about 5 times greater than at Tevatron. This coincides 
with the latest results using PYTHIA \cite{Sjostrand:2004pf,mrenna}. On 
the other hand, the inclusive cross section at LHC is 220 mb according to 
HERWIG, which is about 8 times that at Tevatron. The greater part of the 
difference is presumably due to the difference in the allowed rapidity 
range, i.e., there are more forward scatters at LHC than at Tevatron. We 
have made simple phase-space estimations to confirm this point.

  The HERWIG sub-version 6.505 has appeared recently. One of the major 
ingredients in this release is an interface to JIMMY. It is a simple 
matter to modify this HERWIG--JIMMY package to generate scatters according 
to a Poisson distribution as in our approach. This merely involves the 
substitution of the overlap function by a step function, i.e., 
$A(b^2)=\Theta(\sigma_\mathrm{eff}-\pi b^2)/\sigma_\mathrm{eff}$. We have 
made some calculations by adopting this, and the results are consistent 
with those presented in this study.

 \acknowledgments

  The author thanks Mike Seymour and Bryan Webber for discussions, and Jay
Dittmann, Rick Field, Joey Huston, Mario Martinez-Perez and Ming-Jer Wang
for answering questions related to the details of the CDF charged-jet
study.

 \end{document}